\documentclass[aps,prb,twocolumn,superscriptaddress,longbibliography]{revtex4-2}

\usepackage{physics}
\usepackage[dvipdfmx]{graphicx}
\usepackage{amsmath,amssymb}
\usepackage{xparse}
\usepackage{color}

\definecolor{revisionred}{rgb}{0.85,0.15,0.10}

\def\vk{\vb*{k}}
\def\ii{\mathrm{i}}

\begin{document}

\title{Circular dichroism in second- and third-harmonic generation in the chiral topological semimetal CoSi}

\author{Yuya Ominato}
\email{ominato@aoni.waseda.jp}
\thanks{Contact author}
\affiliation{Waseda Institute for Advanced Study, Waseda University, Shinjuku-ku, Tokyo 169-0051, Japan}
\author{Masahito Mochizuki}
\email{masa\_mochizuki@waseda.jp}
\thanks{Contact author}
\affiliation{Department of Applied Physics, Waseda University, Okubo, Shinjuku-ku, Tokyo 169-8555, Japan}

\date{\today}

\begin{abstract}
We theoretically investigate circular dichroism (CD) in second- and third-harmonic generation (SHG and THG) in the chiral topological semimetal CoSi. We demonstrate that both SHG and THG exhibit dichroic responses of order unity, while their robustness against spectral broadening is strikingly different. Specifically, while SHG-CD is strongly suppressed by dissipation, THG-CD remains robust over a wide frequency range. We show that this qualitative difference originates from the phase structure of the nonlinear current, where SHG-CD arises from subleading interference processes that are sensitive to dephasing, whereas THG-CD emerges already at the leading nonlinear order and is therefore protected against spectral broadening. As a result, THG-CD provides a robust probe of chirality encoded in nonequilibrium electronic dynamics. We further reveal non-monotonic frequency dependences and pronounced sensitivity of harmonic emission to the polarization state and crystallographic orientation of the driving field. Our results provide a representative example of robust nonlinear chiroptical responses in chiral topological semimetals and highlight the potential of high-harmonic spectroscopy as a probe of phase-resolved electronic dynamics.
\end{abstract}

\maketitle 

\section{Introduction}
\label{sec:introduction}

Circular dichroism (CD), namely, the difference in optical response between left-handed and right-handed circularly polarized light, is one of the most fundamental manifestations of chiroptical phenomena \cite{Barron2009,Mun2020,Lininger2023}. In linear optics, CD has long been used as a sensitive probe of chirality in molecules, surfaces, and solids. In crystalline materials, CD is particularly important because it reflects not only the handedness of the lattice structure but also the symmetry and coherence of the electronic states involved in optical transitions \cite{Wang2023,Ocana2023,Zabalo2023,Multunas2023,Urru2025}.
In nonlinear optics, the dichroic response can be much stronger than in ordinary linear CD and can provide microscopic information on the electronic structure and coherence involved in optical processes \cite{Rodrigues2014,Wang2016,Li2017}. Therefore, clarifying the microscopic origin of CD is an important issue in nonlinear chiroptics.

Second- and third-harmonic generation (SHG and THG) are among the most fundamental nonlinear optical responses in solids \cite{Bloembergen1996,Boyd2008}. In noncentrosymmetric crystals, these responses are particularly important because the absence of inversion symmetry allows even-order harmonic generation. From a microscopic point of view, SHG and THG reflect the nonequilibrium electronic dynamics driven by an oscillating electric field. In addition, SHG and THG are sensitive to band structure, quantum coherence, and symmetry constraints. These features make SHG and THG useful probes of noncentrosymmetric electron systems.

These considerations can be extended to SHG and THG driven by circularly polarized light, giving rise to CD in the harmonic response \cite{Petralli1993,Byers1994,Chen2016,Tang2020,Kim2020,Frizyuk2021,Gandolfi2021,Nikitina2023,Toftul2024}. When the emitted harmonic intensity depends on the handedness of the driving field, the resulting dichroic response reflects how crystal symmetry and nonequilibrium electronic dynamics are encoded in nonlinear optical processes \cite{Lovesey2019,Chen2020}. In contrast to linear CD, SHG-CD and THG-CD can involve interference among multiple nonlinear pathways and may therefore exhibit characteristic dependence on the harmonic order, frequency, and dissipation. Thus, clarifying the microscopic origin of such dichroic responses in realistic noncentrosymmetric crystals is an important issue in nonlinear optics.

A particularly suitable platform for addressing these issues is provided by B20-type chiral crystals \cite{Tang2017,Chang2017,Pshenay-Severin2018,Lu2022,Ahn2023,Fan2024}. Among them, CoSi is a representative chiral topological semimetal with space-group 198 symmetry, hosting unconventional band crossings such as a spin-1 chiral fermion at the \(\Gamma\) point and a double Weyl fermion at the R point \cite{Lv2017,Takane2019,Sanchez2019,Schroter2019,Rao2019,Li2019}. These electronic structures, together with the absence of inversion and mirror symmetries, make CoSi an appealing candidate for nonlinear chiroptical responses.

In this work, we theoretically investigate CD in SHG and THG in the chiral topological semimetal CoSi. Using a four-band tight-binding model combined with Floquet theory and the nonequilibrium Green's function method, we show that both SHG and THG exhibit CD of order unity. We find that the frequency dependences of SHG and THG are qualitatively different, and that THG-CD is markedly more robust against spectral broadening than SHG-CD. In SHG, CD arises from the subleading nonlinear interference processes, whereas in THG, CD is already present in the leading nonlinear order. We also show that the harmonic intensity and emission directionality depend strongly on the polarization state and on the crystallographic orientation of the driving field. Our results provide a representative example of CD in nonlinear optical responses arising from the phase structure of harmonic currents and establish high-harmonic spectroscopy as a promising probe of phase-resolved electronic dynamics in chiral quantum materials.

The paper is organized as follows. In Sec.~\ref{sec:model}, we introduce the four-band tight-binding model for CoSi, describe the polarization and orientation of the driving electric field, and present the Floquet-Keldysh formalism for calculating the electric current. In Sec.~\ref{sec:results}, we present the numerical results for the frequency and field-amplitude dependences of SHG and THG under circularly polarized driving, and then examine their dependences on the polarization state and on the crystallographic orientation of the driving field. Sections~\ref{sec:discussion} and \ref{sec:conclusion} are devoted to the discussion and conclusion, respectively.

\begin{figure}[b]
\begin{center}
\includegraphics[width=1\hsize]{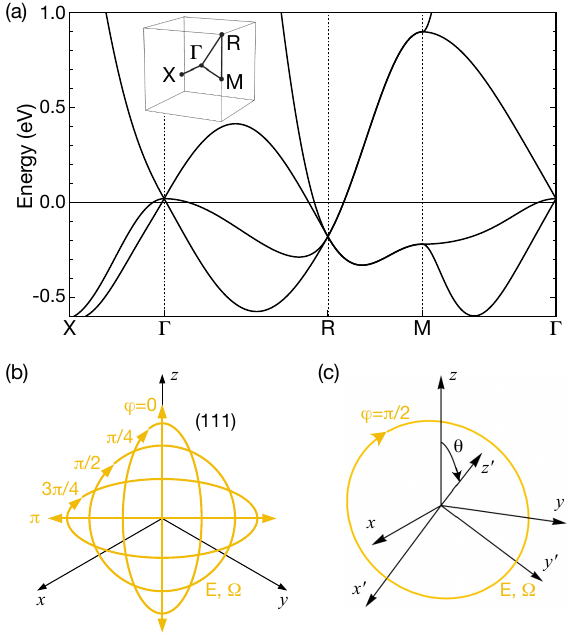}
\end{center}
\caption{
(a) Energy band structure of the tight-binding model without light irradiation.
(b) Light irradiation on the (111) surface. Linear, elliptical, and circular polarizations are considered.
(c) The coordinate system rotated by an angle $\theta$ about the $[\bar{1}10]$ axis is denoted by $x^\prime, y^\prime, z^\prime$. Circularly polarized light is incident along the $z^\prime$ direction, and the electric field rotates within the $x^\prime y^\prime$ plane.}
\label{fig_system}
\end{figure}

\section{Model and Method}
\label{sec:model}

In this section, we present the model and method used to calculate the electric current. We first introduce a four-band tight-binding model for CoSi \cite{Chang2017,Flicker2018}. We then incorporate the light-matter coupling through the Peierls substitution and obtain the corresponding periodically driven Hamiltonian. Finally, we present the formalism used to calculate the electric current based on Floquet theory and the nonequilibrium Green's function method \cite{Tsuji2008,Tsuji2009,Aoki2014,Ominato2025}.

\subsection{Four-band tight-binding model}

We consider a four-band tight-binding model describing the electronic structure near the Fermi level \cite{Chang2017,Flicker2018}. The Hamiltonian is written as
\begin{align}
    H=
    \sum_{\vk}
    (c_{\vk A}^\dagger,c_{\vk B}^\dagger,c_{\vk C}^\dagger,c_{\vk D}^\dagger)
    h(\vk)
    \begin{pmatrix}
    c_{\vk A} \\
    c_{\vk B} \\
    c_{\vk C} \\
    c_{\vk D}
    \end{pmatrix},
\end{align}
where $c_{\vk X}^{\dagger}$ ($c_{\vk X}$) is the creation (annihilation) operator for an electron with crystal momentum $\vk=(k_x,k_y,k_z)$ on sublattice $X=A,B,C,D$.
The crystal structure belongs to space group 198, and the sublattice coordinates within the unit cell are $(0,0,0)$, $(a/2,a/2,0)$, $(a/2,0,a/2)$, and $(0,a/2,a/2)$, where $a=0.44~\mathrm{nm}$ denotes the lattice constant.
The $4\times4$ matrix $h(\vk)$ is given by
\begin{align}
    h(\vk)
    =
    &
        h_{1,xy}(\vk)\mu_0\tau_1
        +
        h_{1,yz}(\vk)\mu_1\tau_0
        +
        h_{1,zx}(\vk)\mu_1\tau_1
     \notag \\
    +
    &
        h_{c,xy}(\vk)\mu_0\tau_2
        +
        h_{c,yz}(\vk)\mu_2\tau_3
        +
        h_{c,zx}(\vk)\mu_2\tau_1
     \notag \\
    +
    &h_2(\vk)I.
    \label{eq_Hk}
\end{align}
The matrix elements are given by
\begin{align}
    &h_{1,ij}(\vk)=4t_1\cos\frac{k_i a}{2}\cos\frac{k_j a}{2}, \\
    &h_{c,ij}(\vk)=4t_c\cos\frac{k_i a}{2}\sin\frac{k_j a}{2}, \\
    &h_2(\vk)=2t_2\sum_{i=x,y,z}\cos k_ia+\varepsilon_0,
\end{align}
where $i,j=x,y,z$, and the parameters are set to $t_1=0.34~\mathrm{eV}$, $t_c=0.14~\mathrm{eV}$, $t_2=0.13~\mathrm{eV}$, and $\varepsilon_0=0.60~\mathrm{eV}$ to reproduce the DFT band structure \cite{Tang2017,Xu2020}.
The matrices are given by
\begin{align}
    &\mu_0\tau_1
    =
    \begin{pmatrix}
        0 & 1 & 0 & 0 \\
        1 & 0 & 0 & 0 \\
        0 & 0 & 0 & 1 \\
        0 & 0 & 1 & 0
    \end{pmatrix},~
    \mu_1\tau_0
    =
    \begin{pmatrix}
        0 & 0 & 1 & 0 \\
        0 & 0 & 0 & 1 \\
        1 & 0 & 0 & 0 \\
        0 & 1 & 0 & 0
    \end{pmatrix},~ \notag \\
    &\mu_1\tau_1
    =
    \begin{pmatrix}
        0 & 0 & 0 & 1 \\
        0 & 0 & 1 & 0 \\
        0 & 1 & 0 & 0 \\
        1 & 0 & 0 & 0
    \end{pmatrix}, 
    \mu_0\tau_2
    =
    \begin{pmatrix}
        0   & -\ii & 0   & 0 \\
        \ii & 0    & 0   & 0 \\
        0   & 0    & 0   & -\ii \\
        0   & 0    & \ii & 0
    \end{pmatrix},~ \notag \\
    &\mu_2\tau_3
    =
    \begin{pmatrix}
        0   & 0    & -\ii & 0 \\
        0   & 0    & 0    & \ii \\
        \ii & 0    & 0    & 0 \\
        0   & -\ii & 0    & 0
    \end{pmatrix},~
    \mu_2\tau_1
    =
    \begin{pmatrix}
        0   & 0   & 0    & -\ii \\
        0   & 0   & -\ii & 0 \\
        0   & \ii & 0    & 0 \\
        \ii & 0   & 0    & 0
    \end{pmatrix},
\end{align}
and $I$ is the $4\times4$ identity matrix.
Diagonalizing Eq.~(\ref{eq_Hk}) along the path shown in Fig.~\ref{fig_system} (a), we obtain the energy band structure.
The spin-1 chiral fermion is observed at the \(\Gamma\) point, and the double Weyl fermion at the R point \cite{Tang2017,Chang2017}.
The energy band structure is consistent with the DFT calculations.

The terms proportional to \(t_c\) break inversion and mirror symmetries, so that the system becomes chiral. The sign of \(t_c\) determines the handedness of the system, and chirality-odd quantities change sign under \(t_c\to -t_c\). The \(\Gamma\) and R points carry topological charges of opposite signs, which are also reversed when the sign of \(t_c\) is inverted.

The four-band Hamiltonian respects space group 198 symmetry. The relevant symmetry operations are the three nonintersecting twofold screw rotations, $s_{2,x}$, $s_{2,y}$, and $s_{2,z}$, together with the threefold rotation $C_{3,[111]}$ about the [111] axis \cite{Bradley1972,Satow2025}. The matrix representations of these operations acting on the sublattice space are given by
\begin{align}
    &s_{2,x}
    =
    \begin{pmatrix}
        0 & 1 & 0 & 0 \\
        1 & 0 & 0 & 0 \\
        0 & 0 & 0 & 1 \\
        0 & 0 & 1 & 0
    \end{pmatrix},
    s_{2,y}
    =
    \begin{pmatrix}
        0 & 0 & 1 & 0 \\
        0 & 0 & 0 & 1 \\
        1 & 0 & 0 & 0 \\
        0 & 1 & 0 & 0
    \end{pmatrix}, \notag \\
    &s_{2,z}
    =
    \begin{pmatrix}
        0 & 0 & 0 & 1 \\
        0 & 0 & 1 & 0 \\
        0 & 1 & 0 & 0 \\
        1 & 0 & 0 & 0
    \end{pmatrix},~
    C_{3,[111]}
    =
    \begin{pmatrix}
        1 & 0 & 0 & 0 \\
        0 & 0 & 0 & 1 \\
        0 & 1 & 0 & 0 \\
        0 & 0 & 1 & 0
    \end{pmatrix}.
\end{align}
Using these representations, the Hamiltonian satisfies
\begin{align}
    &s_{2,i}h(\vk)s_{2,i}^{-1}=h\left(R_{i}(\pi)\vk\right), \\
    &C_{3,[111]}h(\vk)C_{3,[111]}^{-1}=h\left(R_{[111]}(2\pi/3)\vk\right),
\end{align}
where $R_i(\pi)$ denotes the $3\times3$ rotation matrix corresponding to a rotation by $\pi$ about the $i~(=x,y,z)$ axis, and $R_{[111]}(2\pi/3)$ denotes the rotation matrix corresponding to a rotation by $2\pi/3$ about the [111] axis.

\subsection{Periodically driven Hamiltonian}

The vector potential describing the driving electric field is written as
\begin{align}
    \vb*{A}(t)=\frac{E}{\Omega}
    \qty[\sin(\Omega t)\vb*{e}_{x^\prime}+\sin(\Omega t+\varphi)\vb*{e}_{y^\prime}],
\end{align}
where $E$ and $\Omega$ denote the field amplitude and frequency, respectively. The parameter $\varphi$ controls the light polarization, and varying $\varphi$ continuously changes the polarization state, as illustrated in Fig.~\ref{fig_system}(b). The rotated basis vectors $\vb*{e}_{i^\prime}$ are given by
\begin{align}
    \vb*{e}_{i^\prime}=R_{[\bar{1}10]}(\theta)\vb*{e}_i,
\end{align}
where $R_{[\bar{1}10]}(\theta)$ denotes the rotation matrix for a rotation by $\theta$ about the $[\bar{1}10]$ axis, as shown in Fig.~\ref{fig_system}(c). The rotation matrix is explicitly written as
\begin{align}
    R_{[\bar{1}10]}(\theta)
    =\begin{pmatrix}
    \frac{1}{2}(\cos\theta+1) & \frac{1}{2}(\cos\theta-1) & \frac{1}{\sqrt{2}}\sin\theta \\
    \frac{1}{2}(\cos\theta-1) & \frac{1}{2}(\cos\theta+1) & \frac{1}{\sqrt{2}}\sin\theta \\
    -\frac{1}{\sqrt{2}}\sin\theta & -\frac{1}{\sqrt{2}}\sin\theta & \cos\theta
    \end{pmatrix}.
\end{align}
In the $(x,y,z)$ coordinate system, the vector potential is expressed as
\begin{align}
    \vb*{A}(t)
    =
    \frac{E}{\Omega}\sum_{i=x,y,z}
    w_i\sin(\Omega t+\psi_i)\vb*{e}_i,
\end{align}
where
\begin{align}
    &w_i=
    \big[
        (u_i+v_i\cos\varphi)^2
        +
        (v_i\sin\varphi)^2
    \big]^{\frac{1}{2}}, \\
    &\psi_i=
    \mathrm{Arg}\big[
        (u_i+v_i\cos\varphi)
        +
        \ii(v_i\sin\varphi)
    \big],
\end{align}
with
\begin{align}
    &\begin{pmatrix}
        u_x \\
        v_x
    \end{pmatrix}
    =
    \frac{1}{2}
    \begin{pmatrix}
        \cos\theta+1 \\
        \cos\theta-1
    \end{pmatrix}, \\
    &\begin{pmatrix}
        u_y \\
        v_y
    \end{pmatrix}
    =
    \frac{1}{2}
    \begin{pmatrix}
        \cos\theta-1 \\
        \cos\theta+1
    \end{pmatrix}, \\
    &\begin{pmatrix}
        u_z \\
        v_z
    \end{pmatrix}
    =
    -\frac{\sin\theta}{\sqrt{2}}
    \begin{pmatrix}
        1 \\
        1
    \end{pmatrix}.
\end{align}

The periodically driven Hamiltonian is obtained by introducing the light-matter coupling through the Peierls substitution \cite{Peierls1933},
\begin{align}
    h(\vk)\to h\qty(\vk+\frac{e}{\hbar}\vb*{A}(t)).
\end{align}
Since the Hamiltonian is periodic in time, the periodically driven system can be analyzed using Floquet theory \cite{Shirley1965}.
The Hamiltonian matrix in Sambe space is given by \cite{Sambe1973}
\begin{align}
    \mathcal{H}_{ml}(\vk)
    =
    h_{m-l}(\vk)-m\hbar\Omega\delta_{ml}I,
\end{align}
where $h_m(\vk)$ denotes the Fourier component
\begin{align}
    h_m(\vk)
    =
    \int^{T}_0\frac{dt}{T}
    e^{\ii m\Omega t}
    h\qty(\vk+\frac{e}{\hbar}\vb*{A}(t)).
\end{align}
The Fourier component $h_m(\vk)$ is written as
\begin{align}
    &h_m(\vk) \notag \\
    &=
        h_{1,xy,m}(\vk)\mu_0\tau_1
        +
        h_{1,yz,m}(\vk)\mu_1\tau_0
        +
        h_{1,zx,m}(\vk)\mu_1\tau_1
     \notag \\
    &+
        h_{c,xy,m}(\vk)\mu_0\tau_2
        +
        h_{c,yz,m}(\vk)\mu_2\tau_3
        +
        h_{c,zx,m}(\vk)\mu_2\tau_1
     \notag \\
    &+
    h_{2,m}(\vk)\mu_0\tau_0.
\end{align}
The matrix elements are given by
\begin{align}
    &h_{1,ij,m}(\vk) \notag \\
    &=
    t_1
    \sum_{r,s}
    J_{-m}\qty(\mathcal{A}_{1,ij}^{rs})
    e^{\ii(rk_i+sk_j)\frac{a}{2}
    -\ii m\chi_{ij}^{rs}}, \\
    &h_{c,ij,m}(\vk) \notag \\
    &=
    -\ii t_c
    \sum_{r,s}
    sJ_{-m}\qty(\mathcal{A}_{1,ij}^{rs})
    e^{\ii(rk_i+sk_j)\frac{a}{2}
    -\ii m\chi_{ij}^{rs}}, \\
    &h_{2,m}(\vk)=
    t_2
    \sum_{i,r}
    J_{-m}(\mathcal{A}_{2,i}^r)
    e^{\ii rk_i a}
    e^{-\ii m\psi_i}
    +
    \varepsilon_0\delta_{m0},
\end{align}
where $r,s=\pm1$, $i,j=x,y,z$, and $J_{n}(\mathcal{A})$ denote the Bessel functions of the first kind.
Here, the dimensionless quantities $\mathcal{A}_{1,ij}^{rs}$, $\chi_{ij}^{rs}$, and $\mathcal{A}_{2,i}^{r}$ are given by 
\begin{align}
    &\mathcal{A}_{1,ij}^{rs}
    =
    \frac{r}{2}\frac{eEa}{\hbar\Omega}
    \big[(
        w_i\cos\psi_i
        +
        rsw_j\cos\psi_j
    )^2 \notag \\
    &\hspace{2.5cm}+
    (
        w_i\sin\psi_i
        +
        rsw_j\sin\psi_j
    )^2\big]^{\frac{1}{2}}, \\
    &\chi_{ij}^{rs}
    =
    \mathrm{Arg}
    \big[(
        w_i\cos\psi_i
        +
        rsw_j\cos\psi_j
    ) \notag \\
    &\hspace{2.5cm}+
    \ii(
        w_i\sin\psi_i
        +
        rsw_j\sin\psi_j
    )\big], \\
    &\mathcal{A}_{2,i}^{r}
    =
    r\frac{eEa}{\hbar\Omega}w_i.
\end{align}

\subsection{Floquet-Keldysh formalism}

We use the Floquet-Keldysh formalism to describe the time-periodic nonequilibrium state under light irradiation \cite{Tsuji2008,Tsuji2009,Aoki2014}. We assume that such a state is realized through coupling to a reservoir.
The Dyson equation in the Floquet representation is given by
\begin{align}
    &\begin{pmatrix}
        G^{\mathrm{R}}(\vk,\omega) & G^{\mathrm{K}}(\vk,\omega) \\
        0 & G^{\mathrm{A}}(\vk,\omega)
    \end{pmatrix}^{-1} \notag \\
    &=
    \begin{pmatrix}
        G^{0\mathrm{R}}(\vk,\omega) & 0 \\
        0 & G^{0\mathrm{A}}(\vk,\omega)
    \end{pmatrix}^{-1}
    -
    \begin{pmatrix}
        \Sigma^{\mathrm{R}} & \Sigma^{\mathrm{K}}(\omega) \\
        0 & \Sigma^{\mathrm{A}}
    \end{pmatrix},
\end{align}
where the inverses of the unperturbed retarded and advanced Green's functions are given by
\begin{align}
    &\qty[(G^{0\mathrm{R}})^{-1}]_{ml}(\vk,\omega)
    = 
    (\hbar\omega+\ii0)\delta_{ml}I - \mathcal{H}_{ml}(\vk), \\
    &\qty[(G^{0\mathrm{A}})^{-1}]_{ml}(\vk,\omega)
    = 
    (\hbar\omega-\ii0)\delta_{ml}I - \mathcal{H}_{ml}(\vk).
\end{align}
Here, we omit the Keldysh component of the unperturbed Green's function, assuming that the dependence on the initial condition is lost due to dissipation.
The retarded and advanced components of the self-energy are given by
\begin{align}
    &\Sigma^{\mathrm{R}}_{ml}=-\ii\gamma\delta_{ml} I, \\
    &\Sigma^{\mathrm{A}}_{ml}=\ii\gamma\delta_{ml} I,
\end{align}
where \(\gamma\) is a phenomenological parameter characterizing the spectral broadening.
The Keldysh component of the self-energy is given by
\begin{align}
    &\Sigma^{\mathrm{K}}_{ml}(\omega)
    =
    \tanh\qty[\frac{\beta(\hbar\omega+m\hbar\Omega-\mu)}{2}]
    \qty(\Sigma^{\mathrm{R}}_{ml}-\Sigma^{\mathrm{A}}_{ml}),
\end{align}
where $\beta$ is the inverse temperature and $\mu$ is the chemical potential.
This treatment provides a minimal phenomenological description of energy dissipation from the system to the reservoir.

Here, \(\gamma\) should be regarded as a phenomenological parameter representing the combined effects of various relaxation and dephasing processes, rather than a parameter associated with a specific microscopic relaxation mechanism. This treatment provides an effective description of dissipation under periodic driving~\cite{Aoki2014}. In the Floquet-Keldysh formalism used in this work, \(\Sigma^{\mathrm{R/A}}\) determines the spectral broadening, while \(\Sigma^{\mathrm{K}}(\omega)\) determines the occupations and coherences through the Keldysh equation. Similar treatments of dissipation in the Floquet-Keldysh formalism have been shown to reproduce the main features of microscopic quantum kinetic calculations that explicitly include electron-electron and electron-phonon scattering in laser-driven graphene~\cite{Schuler2020}. In addition, our previous work~\cite{Ominato2025} successfully reproduced the frequency dependence of the circular photogalvanic effect in CoSi obtained by perturbative nonlinear-response theory~\cite{Ni2021}. These results support the validity of the present approach as an effective description of dissipation in periodically driven open quantum systems.

To compute the electric current under periodic driving, we need the lesser Green's function. Using the Langreth rules, the lesser Green's function is given by
\begin{align}
    &G^<(\vk,\omega)
    =
    G^{\mathrm{R}}(\vk,\omega)
    \Sigma^<(\omega)
    G^{\mathrm{A}}(\vk,\omega).
    \label{eq_lesser_equation}
\end{align}
The lesser component of the self-energy is related to the retarded, advanced, and Keldysh components as
\begin{align}
    &\Sigma^<(\omega)
    =
    \frac{1}{2}
        \qty[\Sigma^{\mathrm{K}}(\omega)
            -
        \qty(
            \Sigma^{\mathrm{R}}
            -
            \Sigma^{\mathrm{A}}
            )
        ].
\end{align}
The lesser Green's function obtained here is used in the electric current formula below.

The electric current is time periodic under the above assumptions and can be expressed as
\begin{align}
    \expval{\mathcal{J}_i(t)}
    =
    \sum_n e^{-\ii n\Omega t}\expval{\mathcal{J}_{i,n}},
\end{align}
where \(i=x,y,z\). The \(n\)th Fourier component is given by
\begin{align}
    \expval{\mathcal{J}_{i,n}}
    =
    \sum_{\vk}
    \int_{-\Omega/2}^{\Omega/2}\frac{d\omega}{2\pi}
    \mathcal{J}_{i,n}(\vk,\omega), \label{eq_photocurrent}
\end{align}
where the harmonic current spectrum is
\begin{align}
    \mathcal{J}_{i,n}(\vk,\omega)
    =
    2\ii e\hbar\,
    \mathrm{Tr}
    \qty[
        v_i(\vk)S_nG^<(\vk,\omega)
    ].
\end{align}
Here, \(\mathrm{Tr}[\cdot]\) denotes the trace over both the Sambe space and the sublattice indices, \(S_n\) is given by
\begin{align}
    (S_n)_{ml}=\delta_{m,l-n}I,
\end{align}
and the velocity operator is defined as
\begin{align}
    v_i(\vk)
    =
    \frac{1}{\hbar}
    \partial_{k_i}\mathcal{H}(\vk).
\end{align}
The prefactor \(2\) accounts for the spin degeneracy in the present spinless four-band model. In this formulation, we focus on the electric-current response and therefore neglect spin-dependent photoinduced effects. This treatment is justified for the present purpose, since spin-orbit coupling only weakly modifies the electronic band structure of CoSi~\cite{Tang2017}, while the nonlinear optical responses discussed here are primarily governed by the orbital structure of the bands. We note that spin-dependent photoinduced effects under circularly polarized light have been discussed in spin-orbit coupled systems~\cite{Tanaka2020}.
However, such effects lie beyond the scope of the present spinless model and are not expected to qualitatively affect the mechanisms discussed in this work.

The harmonic components of the electric current determine the HG intensity.
The \(n\)th HG intensity is defined as
\begin{align}
    &I_n=I_{\parallel,n}+I_{\perp,n}, \\
    &I_{\parallel,n}=(n\Omega)^2
    \qty(
        \abs{\expval{\mathcal{J}_{{x^\prime},n}}}^2
        +
        \abs{\expval{\mathcal{J}_{{y^\prime},n}}}^2
        ), \\
    &I_{\perp,n}=(n\Omega)^2\abs{\expval{\mathcal{J}_{{z^\prime},n}}}^2.
\end{align}
The HG intensity is decomposed into components parallel and perpendicular to the plane in which the driving electric field oscillates. To quantify circular dichroism in $n$th HG, we introduce the dimensionless dichroism factor
\begin{align}
    g_{n}^{\mathrm{CD}}
    =
    \frac
    {I_n^{\circlearrowleft}-I_n^{\circlearrowright}}
    {\frac{1}{2}(I_n^{\circlearrowleft}+I_n^{\circlearrowright})},
\end{align}
where $I_n^{\circlearrowleft}$ and $I_n^{\circlearrowright}$ denote the intensities for $\varphi=-\pi/2$ and $\varphi=\pi/2$, respectively, corresponding to left-handed and right-handed circularly polarized light.

An advantage of the Floquet-Keldysh formalism for the present purpose is that the electric current can be evaluated without an explicit order-by-order expansion in the field amplitude \(E\). This allows us to include higher-order nonlinear contributions within a unified framework without separately evaluating the complex expressions containing many matrix elements and energy denominators that arise in an order-by-order perturbative analysis \cite{Morimoto2016}.

\section{Results}
\label{sec:results}

The results presented below are obtained under the following common conditions. We set \(\mu=0\) in the main text, while results for \(\mu\neq0\) are given in the Appendix, where we confirm that the main conclusions remain qualitatively unchanged. The inverse temperature is fixed at \(\beta^{-1}=0.03~\mathrm{eV}\). We have confirmed that changing \(\beta^{-1}\) by an order of magnitude leads to only minor quantitative changes and does not affect the conclusions. The photon number is restricted to \(|m|\le 6\) in all numerical calculations, and Eq.~(\ref{eq_photocurrent}) is discretized and numerically integrated over crystal momentum \(\vk\) and frequency \(\omega\) using a \(60^3\times 40\) mesh. Under these conditions, the numerical results are sufficiently converged, and further increasing the photon-number cutoff or the mesh resolution leads to only minor quantitative changes.

\begin{figure*}[t]
\begin{center}
\includegraphics[width=1\hsize]{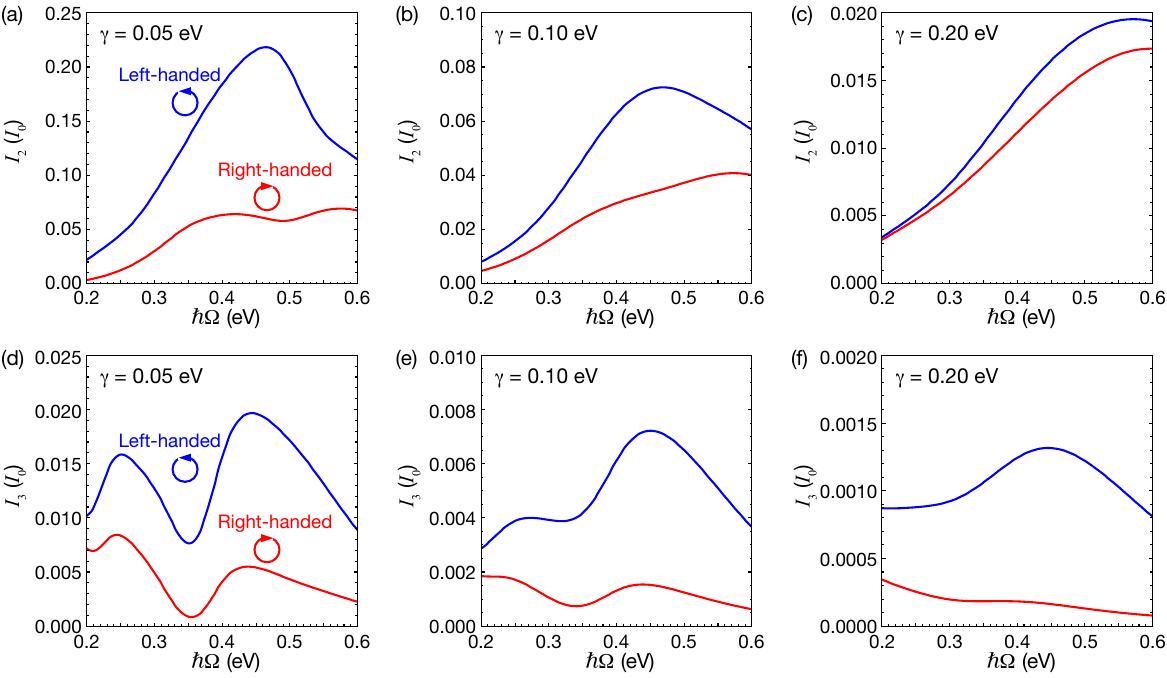}
\end{center}
\caption{SHG intensity $I_2$ as a function of $\hbar\Omega$ for (a) $\gamma=0.05$ eV, (b) $\gamma=0.10$ eV, and (c) $\gamma=0.20$ eV. THG intensity $I_3$ as a function of $\hbar\Omega$ for (d) $\gamma=0.05$ eV, (e) $\gamma=0.10$ eV, and (f) $\gamma=0.20$ eV. 
The unit is given by $I_0=\qty({2e^3\mathrm{V}^2}/{(16\pi^4\hbar^2 a^2)})^2\simeq 6\times10^6\mathrm{A^2/(\mu m^4ps^2)}$.
The blue and red curves represent the intensities for \(\varphi=-\pi/2\) and \(\varphi=\pi/2\), respectively, corresponding to left-handed and right-handed circularly polarized light.
We set $E = 2.0~\mathrm{MV/cm}$ and $\beta^{-1} = 0.03~\mathrm{eV}$.
}
\label{fig_CD_w}
\end{figure*}

\subsection{Frequency dependence of the SHG and THG under circularly polarized light in the (111) plane}

In this subsection, we consider circularly polarized light whose electric field rotates in the (111) plane. According to the selection rules for HHG \cite{Simon1968,Tang1971,Alon1998,Saito2017,Neufeld2019,Ikeda2019,Ikeda2020,Kanega2021,Kanega2024,Ominato2025}, the SHG intensity has only an in-plane component $(I_{\parallel,2}\neq0,\ I_{\perp,2}=0)$, whereas the THG intensity has only an out-of-plane component $(I_{\parallel,3}=0,\ I_{\perp,3}\neq0)$.

Figure \ref{fig_CD_w} shows the frequency dependence of the SHG and THG intensities under left-handed and right-handed circularly polarized light, revealing CD in both cases.
As seen in Fig.~\ref{fig_CD_w}(a), the SHG intensities for left-handed and right-handed circularly polarized light are clearly different over the entire frequency range, indicating a pronounced SHG-CD.
For left-handed circularly polarized light, the SHG intensity takes a maximum slightly below \(\hbar\Omega=0.5~\mathrm{eV}\). In contrast, for right-handed circularly polarized light, multiple local maxima are observed, together with a shallow dip around \(\hbar\Omega=0.5~\mathrm{eV}\). Panels (a)--(c) show the results calculated for different values of \(\gamma\). A comparison of these panels shows that the SHG intensity tends to decrease as \(\gamma\) increases. At the same time, the difference between left-handed and right-handed circularly polarized light is also reduced. In addition, the frequency dependence becomes more gradual.

As seen in Fig.~\ref{fig_CD_w}(d), the THG intensities for left-handed and right-handed circularly polarized light are also clearly different over the entire frequency range, indicating a pronounced THG-CD. Figure~\ref{fig_CD_w}(d) also shows a clear dip around \(\hbar\Omega=0.35~\mathrm{eV}\).
The dip can be understood from the phase structure of the harmonic current spectrum, as explained later in detail.
Panels (d)--(f) show the results calculated for different values of \(\gamma\).
A comparison of these panels shows that the THG intensity tends to decrease and the dip is suppressed as \(\gamma\) increases.
The THG intensity is about one order of magnitude smaller than the SHG intensity for the present parameter set.
A notable difference from SHG is that the difference between the THG intensities for left-handed and right-handed circularly polarized light remains relatively insensitive to increasing \(\gamma\).

\begin{figure}[t]
\begin{center}
\includegraphics[width=1\hsize]{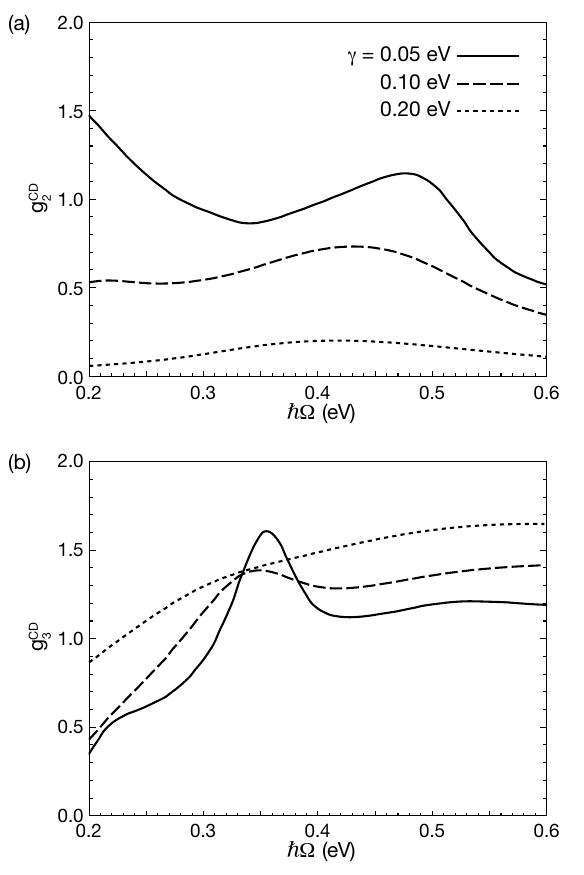}
\end{center}
\caption{(a) \(g_2^{\mathrm{CD}}\) and (b) \(g_3^{\mathrm{CD}}\) as functions of \(\hbar\Omega\) for \(\gamma=0.05~\mathrm{eV}\), \(0.10~\mathrm{eV}\), and \(0.20~\mathrm{eV}\). We set \(E=2.0~\mathrm{MV/cm}\) and \(\beta^{-1}=0.03~\mathrm{eV}\).
}
\label{fig_CD_g}
\end{figure}

Figure~\ref{fig_CD_g} shows the frequency dependence of \(g_2^{\mathrm{CD}}\) and \(g_3^{\mathrm{CD}}\). As seen in Fig.~\ref{fig_CD_g}(a), SHG-CD is suppressed over the entire frequency range as \(\gamma\) increases. For \(\gamma=0.05~\mathrm{eV}\), SHG-CD exhibits a local maximum near \(\hbar\Omega=0.5~\mathrm{eV}\), where the SHG intensity for left-handed circularly polarized light also takes a local maximum. In contrast, Fig.~\ref{fig_CD_g}(b) shows that THG-CD remains of order unity over the entire frequency range, and its magnitude tends to increase with increasing \(\gamma\), except near \(\hbar\Omega=0.35~\mathrm{eV}\). For \(\gamma=0.05~\mathrm{eV}\), THG-CD exhibits a local maximum near \(\hbar\Omega=0.35~\mathrm{eV}\), where the THG intensity shows the dip, as seen in Fig.~\ref{fig_CD_w}(d). The frequency dependence of both SHG-CD and THG-CD becomes less pronounced as \(\gamma\) increases.

\begin{figure}[t]
\begin{center}
\includegraphics[width=1\hsize]{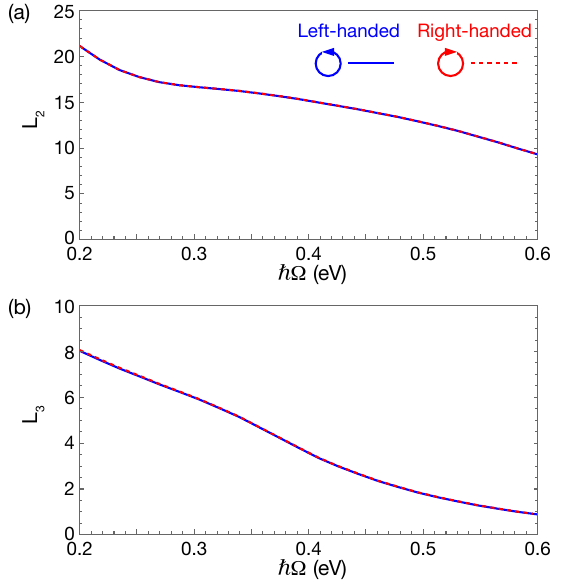}
\end{center}
\caption{
(a) \(L_2\) and (b) \(L_3\) as functions of \(\hbar\Omega\).
The blue solid and red dashed curves correspond to \(\varphi=-\pi/2\) and \(\varphi=\pi/2\), respectively.
We set \(E=2.0~\mathrm{MV/cm}\), \(\gamma=0.05~\mathrm{eV}\), and \(\beta^{-1}=0.03~\mathrm{eV}\).
}
\label{fig_CD_a}
\end{figure}

The emergence of SHG-CD and THG-CD, as well as the non-monotonic frequency dependence of the SHG and THG intensities, can be traced back to the phase structure of the harmonic current spectrum \(\mathcal{J}_{i,n}(\vk,\omega)\), which can be written in terms of its magnitude and phase as
\begin{align}
    \mathcal{J}_{i,n}(\vk,\omega)
    =
    \abs{\mathcal{J}_{i,n}(\vk,\omega)}e^{\ii \vartheta_{i,n}(\vk,\omega)}.
\end{align}
The phase \(\vartheta_{i,n}(\vk,\omega)\) is determined by the Fourier components of the velocity operator and the lesser Green's function.
To clarify the role of the phase structure, we evaluate the frequency dependence of the integrated magnitude of the harmonic current spectrum,
\begin{align}
    L_{n}
    =
    \sum_{\vk}
    \int_{-\Omega/2}^{\Omega/2}\frac{d\omega}{2\pi}
    \abs{\mathcal{J}_{x,n}(\vk,\omega)}.
    \label{eq_indicator}
\end{align}
Figures~\ref{fig_CD_a}(a) and \ref{fig_CD_a}(b) show \(L_2\) and \(L_3\), respectively, for left-handed and right-handed circularly polarized light. As shown in these panels, the difference between the results for left-handed and right-handed circularly polarized light is negligible.
Moreover, even for \(\gamma=0.05~\mathrm{eV}\), the curves vary monotonically as a function of \(\hbar\Omega\), in sharp contrast to the corresponding SHG and THG intensities.
These results indicate that the phase structure of the harmonic current spectrum plays a decisive role in the circular dichroism and the non-monotonic frequency dependence.

To gain microscopic insight into the phase structure, we decompose the harmonic current spectrum into intraband and interband contributions by transforming the velocity operator in the Floquet-state basis.
The Floquet Hamiltonian is diagonalized by a unitary matrix \(\mathcal{U}(\vk)\) as
\begin{align}
    \mathcal{H}(\vk)
    =
    \mathcal{U}(\vk)
    \mathcal{E}(\vk)
    \mathcal{U}^\dagger(\vk),
\end{align}
where \(\mathcal{E}(\vk)\) is the diagonal matrix of quasienergies and the columns of \(\mathcal{U}(\vk)\) correspond to the Floquet states. The same unitary transformation diagonalizes the retarded and advanced Green's functions as
\begin{align}
    G^{\mathrm{R}/\mathrm{A}}(\vk,\omega)
    =
    \mathcal{U}(\vk)
    g^{\mathrm{R}/\mathrm{A}}(\vk,\omega)
    \mathcal{U}^\dagger(\vk).
\end{align}
In this basis, the velocity operator is decomposed into the group-velocity term and the geometric correction term as
\begin{align}
    v_i(\vk)
    =
    \mathcal{U}(\vk)
    \qty(
        \frac{1}{\hbar}
        \partial_{k_i}\mathcal{E}(\vk)
        +
        \frac{\ii}{\hbar}[\xi_i(\vk),\mathcal{E}(\vk)]
    )
    \mathcal{U}^\dagger(\vk),
\end{align}
where \(\xi_i(\vk)\) is the non-Abelian Berry connection defined by
\begin{align}
    \xi_i(\vk)
    =
    -\ii\,\mathcal{U}^\dagger(\vk)
    \partial_{k_i}\mathcal{U}(\vk).
\end{align}
Substituting these expressions into the trace, we obtain
\begin{align}
    \mathcal{J}_{i,n}(\vk,\omega)
    =
    \mathcal{J}_{i,n}^{\mathrm{intra}}(\vk,\omega)
    +
    \mathcal{J}_{i,n}^{\mathrm{inter}}(\vk,\omega),
\end{align}
where the two contributions are given by
\begin{align}
    &\mathcal{J}_{i,n}^{\mathrm{intra}}(\vk,\omega) \notag \\
    &=
    2\ii e\,
    \mathrm{Tr}
    \qty[
        \partial_{k_i}\mathcal{E}(\vk)
        S_n
        g^{\mathrm{R}}(\vk,\omega)
        \tilde{\Sigma}^<(\vk,\omega)
        g^{\mathrm{A}}(\vk,\omega)
    ], \\
    &\mathcal{J}_{i,n}^{\mathrm{inter}}(\vk,\omega) \notag \\
    &=
    -2e\,
    \mathrm{Tr}
    \qty[
        [\xi_i(\vk),\mathcal{E}(\vk)]
        S_n
        g^{\mathrm{R}}(\vk,\omega)
        \tilde{\Sigma}^<(\vk,\omega)
        g^{\mathrm{A}}(\vk,\omega)
    ].
\end{align}
Here, we have introduced
\begin{align}
    \tilde{\Sigma}^<(\vk,\omega)
    =
    \mathcal{U}^\dagger(\vk)
    \Sigma^<(\vk,\omega)
    \mathcal{U}(\vk),
\end{align}
which denotes the lesser self-energy in the Floquet-state basis. The intraband contribution originates from the diagonal group velocity \(\hbar^{-1}\,\partial_{k_i}\mathcal{E}(\vk)\), while the interband contribution originates from the off-diagonal geometric term \(\ii\hbar^{-1}\,[\xi_i(\vk),\mathcal{E}(\vk)]\). In the present calculation, both contributions are included in the harmonic current spectrum. Moreover, since \(\tilde{\Sigma}^<(\vk,\omega)\) is generally not diagonal in the Floquet-state basis, the lesser self-energy determines not only the Floquet-state populations but also the coherences between different Floquet states through Eq.~(\ref{eq_lesser_equation}). As a result, the phase of the nonlinear current reflects the micromotion of the Floquet states, the geometric correction to the velocity operator, and the dissipation-induced mixing between different Floquet states. From this perspective, the dip structure, SHG-CD, and THG-CD can be understood as interference phenomena governed by the resulting phase structure \(\vartheta_{i,n}(\vk,\omega)\) of the nonlinear harmonic current.

\subsection{Electric-field dependence of the SHG and THG under circularly polarized light in the (111) plane}

\begin{figure*}[t]
\begin{center}
\includegraphics[width=1\hsize]{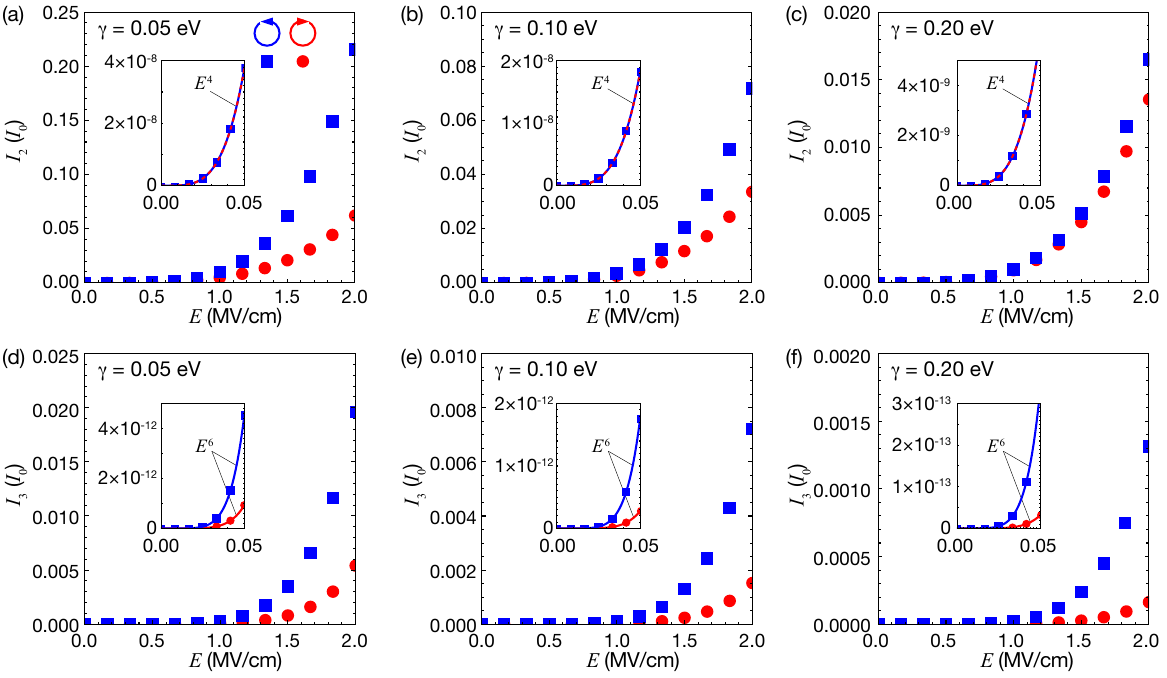}
\end{center}
\caption{SHG intensity $I_2$ as a function of $E$ for (a) $\gamma=0.05~\mathrm{eV}$, (b) $\gamma=0.10~\mathrm{eV}$, and (c) $\gamma=0.20~\mathrm{eV}$. THG intensity $I_3$ as a function of $E$ for (d) $\gamma=0.05~\mathrm{eV}$, (e) $\gamma=0.10~\mathrm{eV}$, and (f) $\gamma=0.20~\mathrm{eV}$.
The unit is given by $I_0\simeq 6\times10^6\mathrm{A^2/(\mu m^4ps^2)}$.
The blue square and red circle markers represent the intensities for $\varphi=-\pi/2$ and $\pi/2$, respectively. The blue and red curves represent fits in the weak-field region for $\varphi=-\pi/2$ and $\varphi=\pi/2$, respectively. The fitted curves are proportional to $E^4$ for panels (a)--(c) and to $E^6$ for panels (d)--(f). We set \(\beta^{-1} = 0.03~\mathrm{eV}\) and \(\hbar\Omega = 0.45~\mathrm{eV}\).
}
\label{fig_CD_e}
\end{figure*}

Figure~\ref{fig_CD_e} shows the electric-field dependence of the SHG and THG intensities at \(\hbar\Omega=0.45~\mathrm{eV}\). Panels (a)--(c) present the SHG intensities, and panels (d)--(f) show the THG intensities. We first discuss the SHG intensity in Fig.~\ref{fig_CD_e}(a). The difference between the SHG intensities for left-handed and right-handed circularly polarized light becomes more pronounced as the electric field increases. The inset shows the weak-field region. The markers represent the numerical data, and the curves in the inset show \(E^4\) fits to the weak-field data. In the weak-field region, the numerical data are well reproduced by the \(E^4\) fit, and the difference between the SHG intensities for left-handed and right-handed circularly polarized light is negligible. These results indicate that the circular dichroism arises from higher-order nonlinear contributions beyond the leading \(E^4\) contribution.

A comparison of Figs.~\ref{fig_CD_e}(a)--(c) shows the \(\gamma\) dependence of the SHG intensity. As already shown in Fig.~\ref{fig_CD_w}, the difference between the SHG intensities for left-handed and right-handed circularly polarized light is suppressed as \(\gamma\) increases. At the same time, the insets show that the \(E^4\) fit remains valid in the weak-field region for all values of \(\gamma\). These results suggest that higher-order nonlinear contributions beyond the leading \(E^4\) contribution are more strongly suppressed as \(\gamma\) increases, leading to the reduction of SHG-CD.

Figure~\ref{fig_CD_e}(d) shows the THG intensity. As in the SHG case, a clear difference between the THG intensities for left-handed and right-handed circularly polarized light is observed. There is, however, an important difference from SHG. As shown in the inset for the weak-field region, the numerical data are well reproduced by an \(E^6\) fit, and the difference between left-handed and right-handed circularly polarized light is already present in the leading nonlinear contribution.

A comparison of Figs.~\ref{fig_CD_e}(d)--(f) shows the \(\gamma\) dependence of the THG intensity. Although the overall THG intensity decreases as \(\gamma\) increases, the qualitative features remain unchanged. In particular, the \(E^6\) fit remains valid in the weak-field region, and the difference between the THG intensities for left-handed and right-handed circularly polarized light is present for all values of \(\gamma\). This behavior contrasts with SHG, where circular dichroism arises beyond the leading order nonlinear contribution. This difference in the leading nonlinear response is reflected in the distinct \(\gamma\) dependences of SHG-CD and THG-CD.

\subsection{Polarization dependence in the (111) plane}

\begin{figure}[!t]
\begin{center}
\includegraphics[width=1\hsize]{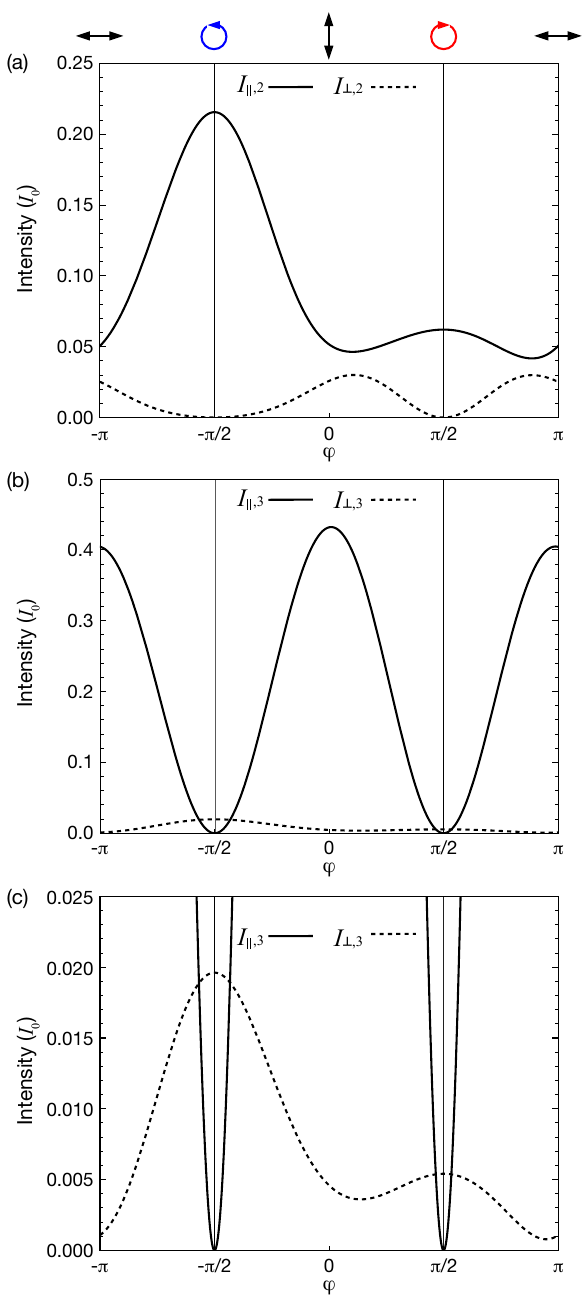}
\end{center}
\caption{
(a) SHG intensity components \(I_{\parallel,2}\) and \(I_{\perp,2}\) as functions of \(\varphi\). (b), (c) THG intensity components \(I_{\parallel,3}\) and \(I_{\perp,3}\) as functions of \(\varphi\), shown on different intensity scales. The unit is \(I_0\simeq 6\times10^6~\mathrm{A^2/(\mu m^4 ps^2)}\). We set \(E=2.0~\mathrm{MV/cm}\), \(\gamma=0.05~\mathrm{eV}\), \(\beta^{-1}=0.03~\mathrm{eV}\), and \(\hbar\Omega = 0.45~\mathrm{eV}\).
}
\label{fig_CD_f}
\end{figure}

In this subsection, we show the polarization dependence of the SHG and THG intensities for an electric field oscillating in the (111) plane. As discussed in the previous subsection, the selection rules for circular polarization require that only the in-plane component can be finite for SHG, whereas only the out-of-plane component can be finite for THG. Once the phase difference \(\varphi\) is varied away from \(\pm\pi/2\), both the in-plane and out-of-plane components become finite for SHG and THG.

Figure~\ref{fig_CD_f}(a) shows the polarization dependence of the SHG intensity. The arrows shown above the panel schematically indicate the polarization states. Over the entire range of \(\varphi\), the in-plane component is larger than the out-of-plane component. The in-plane component takes its local maximum at \(\varphi=\pm\pi/2\), and the out-of-plane component becomes finite as \(\varphi\) deviates from $\pm\pi/2$.

Figures~\ref{fig_CD_f}(b) and \ref{fig_CD_f}(c) show the polarization dependence of the THG intensity.
Figure~\ref{fig_CD_f}(c) shows the same THG data on a reduced vertical scale so that the smaller component can be seen more clearly.
In contrast to SHG, the out-of-plane component is dominant near \(\varphi=\pm\pi/2\), because the in-plane component is forbidden at \(\varphi=\pm\pi/2\) by the selection rule for circular polarization. As \(\varphi\) deviates from \(\pm\pi/2\), the in-plane component grows and becomes dominant. Thus, while SHG is dominated by the in-plane response for all polarization states, THG changes from an out-of-plane-dominant response near \(\varphi=\pm\pi/2\) to an in-plane-dominant response as \(\varphi\) moves away from \(\pm\pi/2\).

\subsection{Crystallographic-orientation dependence}

\begin{figure}[!t]
\begin{center}
\includegraphics[width=1\hsize]{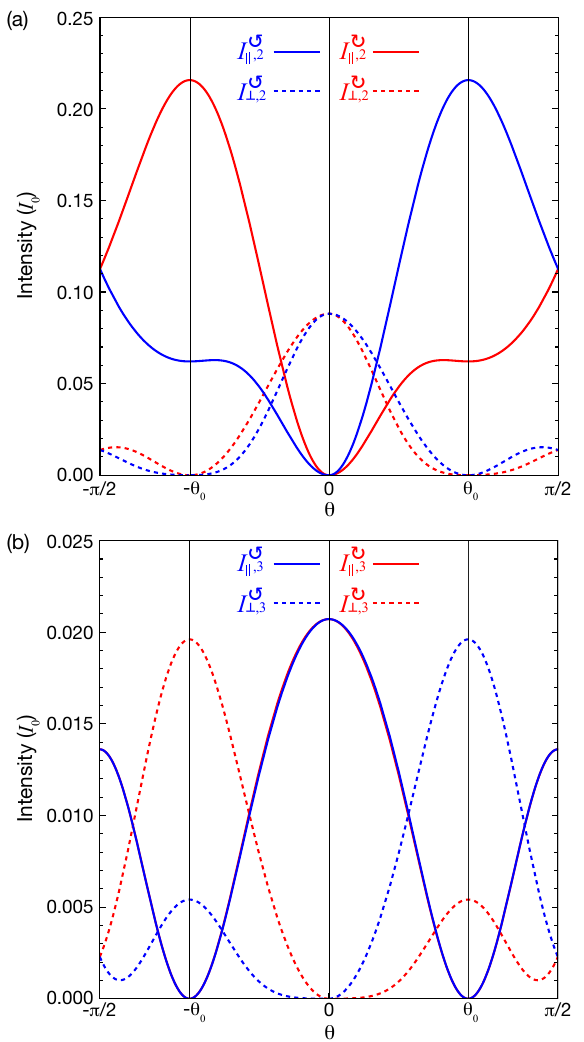}
\end{center}
\caption{
(a) SHG intensities $I_{\parallel,2}^{\circlearrowleft/\circlearrowright}$ and $I_{\perp,2}^{\circlearrowleft/\circlearrowright}$ as functions of $\theta$.
(b) THG intensities $I_{\parallel,3}^{\circlearrowleft/\circlearrowright}$ and $I_{\perp,3}^{\circlearrowleft/\circlearrowright}$ as functions of $\theta$.
The unit is given by $I_0\simeq 6\times10^6\mathrm{A^2/(\mu m^4ps^2)}$.
We set $E=2.0~\mathrm{MV/cm}$, $\gamma=0.05~\mathrm{eV}$, $\beta^{-1}=0.03~\mathrm{eV}$, and \(\hbar\Omega = 0.45~\mathrm{eV}\).
}
\label{fig_CD_q}
\end{figure}

In this subsection, we show the SHG and THG intensities for circular polarization when the rotation plane of the driving electric field is tilted away from the (111) plane, as illustrated in Fig.~\ref{fig_system}(c). Figures~\ref{fig_CD_q}(a) and \ref{fig_CD_q}(b) show the SHG and THG intensities, respectively, resolved into their in-plane and out-of-plane components. The blue and red curves correspond to \(\varphi=-\pi/2\) and \(\varphi=\pi/2\), respectively, corresponding to left-handed and right-handed circularly polarized light. Here, \(\theta_0\) denotes the angle at which the \(z^\prime\) axis coincides with the [111] axis. The results are consistent with the selection rules \cite{Simon1968,Tang1971,Alon1998,Saito2017,Neufeld2019,Ikeda2019,Ikeda2020,Kanega2021,Kanega2024,Ominato2025}. In particular, the circular dichroism vanishes at \(\theta=0\) and \(\pm\pi/2\). In addition, at \(\theta=\pm\theta_0\), the out-of-plane component is forbidden for SHG, whereas the in-plane component is forbidden for THG. The results for left-handed and right-handed circularly polarized light are mapped onto each other by \(\theta\to-\theta\), as expected from the crystal symmetry.

As shown in Fig.~\ref{fig_CD_q}(a), only the out-of-plane component of SHG is finite at \(\theta=0\). As \(\theta\) deviates from \(0\), the in-plane component grows and eventually becomes larger than the out-of-plane component. Beyond this crossing, the in-plane component remains dominant up to \(\theta=\pm\pi/2\). Thus, for SHG, the dominant response changes from out-of-plane near \(\theta=0\) to in-plane away from \(\theta=0\).

As shown in Fig.~\ref{fig_CD_q}(b), the in-plane component of THG is finite and the out-of-plane component vanishes at \(\theta=0\). As \(\theta\) deviates from \(0\), the in-plane component decreases and vanishes at \(\theta=\pm\theta_0\), while the out-of-plane component increases and reaches its local maximum there. Beyond \(\theta=\pm\theta_0\), the in-plane component grows and eventually becomes larger than the out-of-plane component near \(\theta=\pm\pi/2\). Therefore, the in-plane and out-of-plane responses show qualitatively different \(\theta\) dependences in SHG and THG.

\section{Discussion}
\label{sec:discussion}

It is worth mentioning the relation between the present results and the chirality of the system. Because the analysis in this work focuses on the harmonic intensities, the main quantities discussed here are chirality even. In other words, the quantities plotted in the main text remain unchanged when the chirality of the system is reversed, namely, under \(t_c\to -t_c\). In this sense, the order-unity SHG-CD and THG-CD reported here do not directly probe the handedness of CoSi, but primarily reflect the noncentrosymmetric nature of the system.

In contrast, the Fourier components of the electric current retain chirality-odd information. Although the corresponding data are not shown, we have confirmed that, under \(t_c\to -t_c\), the Fourier components of the electric current transform as
\begin{align}
    \expval{\mathcal{J}_{i,n}}
    \to
    \begin{cases}
        -\expval{\mathcal{J}_{i,n}} & (n: \ \text{even}) \\
        \phantom{-}\expval{\mathcal{J}_{i,n}} & (n: \ \text{odd})
    \end{cases}.
\end{align}
In particular, the \(n=0\) component is chirality odd, which is consistent with the sign reversal of the dc photocurrent under reversal of the system handedness \cite{Rees2020,Ni2020,Ni2021}. These results suggest that chirality-odd properties in higher-harmonic responses can be accessed only through observables that retain the phase information of \(\expval{\mathcal{J}_{i,n}}\), such as real-time electric current dynamics or phase-sensitive measurements of electric current. A more detailed investigation of this issue is left for future work.

\section{Conclusion}
\label{sec:conclusion}

In this work, we have theoretically investigated circular dichroism in second- and third-harmonic generation in the chiral topological semimetal CoSi.
We find that both SHG and THG exhibit circular dichroism of order unity, while their frequency dependences are qualitatively different.
In addition, THG-CD is markedly more robust against spectral broadening than SHG-CD. Our analysis of the harmonic current spectrum shows that the non-monotonic frequency dependence of the harmonic intensities, as well as the circular dichroism, can be traced back to the phase structure of the harmonic current spectrum. SHG-CD arises from higher-order nonlinear contributions beyond the leading nonlinear response, whereas THG-CD is already present in the leading nonlinear response. We also clarified how the harmonic-generation intensities depend on the polarization state and on the crystallographic orientation. Our findings provide a representative example of a robust nonlinear chiroptical response in chiral topological semimetals and demonstrate the potential of high-harmonic spectroscopy as a probe of phase-resolved electronic dynamics.

\section*{Acknowledgments}

This work was supported by JSPS KAKENHI (Grants No.~JP24H02231, No.~JP25K17939, No.~JP25H00611, No.~JP25KF0118, and No.~JP26K06985), JST CREST (Grant No.~JPMJCR20T1), and Waseda University Grant for Special Research Projects (Grants No.~2025C-133, No.~2025C-651, No.~2025R-061, and No.~2026C-668).

\appendix

\section{Chemical-potential dependence of SHG-CD and THG-CD in the (111) plane}

\begin{figure}[t]
\begin{center}
\includegraphics[width=1\hsize]{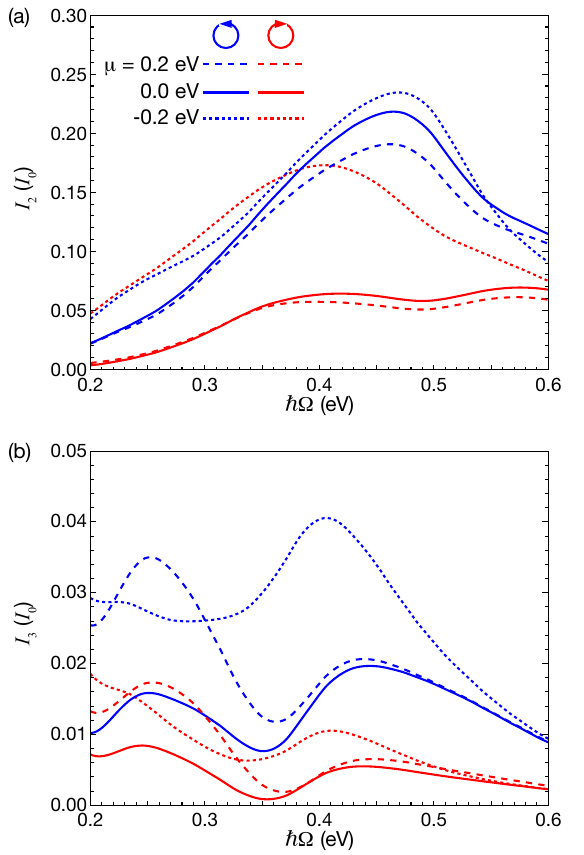}
\end{center}
\caption{
(a) SHG intensities \(I_2\) and (b) THG intensities \(I_3\) as functions of \(\hbar\Omega\) for several values of \(\mu=-0.2~\mathrm{eV}\), \(0.0~\mathrm{eV}\), and \(0.2~\mathrm{eV}\).
The unit is given by \(I_0\simeq 6\times10^6\mathrm{A^2/(\mu m^4ps^2)}\).
The blue and red curves represent the intensities for \(\varphi=-\pi/2\) and \(\pi/2\), respectively, corresponding to left-handed and right-handed circularly polarized light. We set \(E = 2.0~\mathrm{MV/cm}\), \(\gamma = 0.05~\mathrm{eV}\), and \(\beta^{-1} = 0.03~\mathrm{eV}\).
}
\label{fig_CD_app1}
\end{figure}

\begin{figure}[t]
\begin{center}
\includegraphics[width=1\hsize]{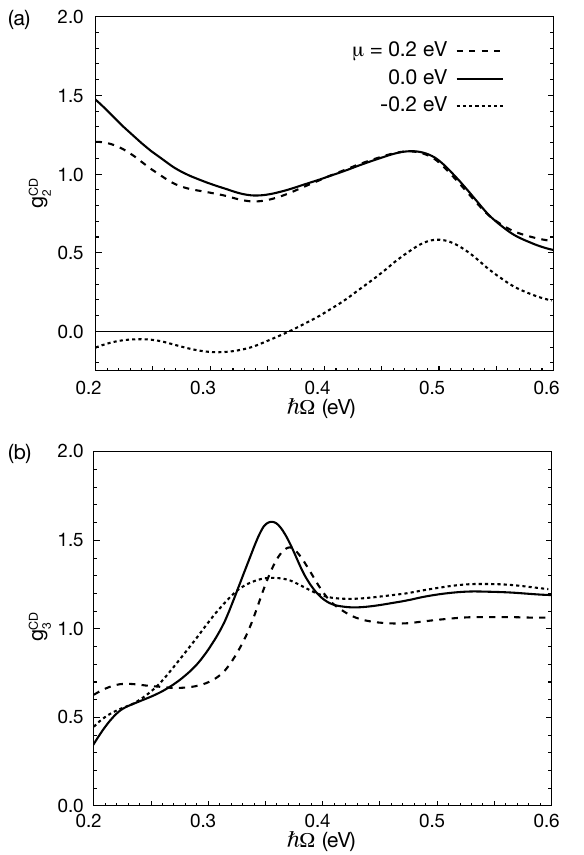}
\end{center}
\caption{
(a) SHG-CD \(g_2^{\mathrm{CD}}\) and (b) THG-CD \(g_3^{\mathrm{CD}}\) as functions of \(\hbar\Omega\) for \(\mu=-0.2~\mathrm{eV}\), \(0.0~\mathrm{eV}\), and \(0.2~\mathrm{eV}\). We set \(E=2.0~\mathrm{MV/cm}\), \(\gamma=0.05~\mathrm{eV}\), and \(\beta^{-1}=0.03~\mathrm{eV}\).
}
\label{fig_CD_app2}
\end{figure}

In this Appendix, we examine the chemical-potential dependence of SHG-CD and THG-CD on the (111) surface. Overall, both SHG-CD and THG-CD remain finite when \(\mu\) is shifted away from \(0\), and their qualitative behaviors are largely preserved. An exception is found for SHG-CD at \(\mu=-0.2~\mathrm{eV}\). In this case, the difference between the SHG intensities for left-handed and right-handed circularly polarized light is reduced, and the sign of \(g^{\mathrm{CD}}_{2}\) reverses in the low-frequency region.

Figure~\ref{fig_CD_app1} shows the SHG intensity in panel (a) and the THG intensity in panel (b) as functions of \(\hbar\Omega\) for \(\mu=-0.2~\mathrm{eV}\), \(0.0~\mathrm{eV}\), and \(0.2~\mathrm{eV}\). We first discuss the SHG results in panel (a). The blue and red curves represent the intensities for \(\varphi=-\pi/2\) and \(\varphi=\pi/2\), respectively, corresponding to left-handed and right-handed circularly polarized light. Among the blue curves, changing \(\mu\) leads to only minor quantitative changes. In contrast, the red curves show a more pronounced dependence on \(\mu\). For \(\mu=-0.2~\mathrm{eV}\), the intensity is substantially modified, and the difference between the SHG intensities for left-handed and right-handed circularly polarized light is reversed in the low-frequency region. We next discuss the THG results in panel (b). For all chemical potentials considered, the THG intensity for left-handed circularly polarized light remains larger than that for right-handed circularly polarized light.
The dependence of the SHG and THG intensities on \(\mu\) is more pronounced in the low-frequency region than in the high-frequency region. Since the nodal points are located close to the Fermi level, this enhanced sensitivity suggests a close connection with the topological and quantum-geometric properties of the low-energy multifold fermions in CoSi.

Figure~\ref{fig_CD_app2} shows the corresponding SHG-CD in panel (a) and THG-CD in panel (b) as functions of \(\hbar\Omega\) for \(\mu=-0.2~\mathrm{eV}\), \(0.0~\mathrm{eV}\), and \(0.2~\mathrm{eV}\). As seen in panel (a), \(g^{\mathrm{CD}}_{2}\) becomes negative in the low-frequency region for \(\mu=-0.2~\mathrm{eV}\), and its magnitude is also reduced. Aside from this case, both SHG-CD and THG-CD show only minor quantitative changes, while their overall magnitudes and trends are largely preserved.

%\clearpage
\bibliography{ref}

@article{Tanaka2020,
  title={Theory of the inverse Faraday effect due to the Rashba spin--orbit interactions: roles of band dispersions and Fermi surfaces},
  author={Tanaka, Yasuhiro and Inoue, Takashi and Mochizuki, Masahito},
  journal={New Journal of Physics},
  volume={22},
  number={8},
  pages={083054},
  year={2020},
  publisher={IOP Publishing}
}

@article{Schuler2020,
  title = {How Circular Dichroism in Time- and Angle-Resolved Photoemission Can Be Used to Spectroscopically Detect Transient Topological States in Graphene},
  author = {Sch\"uler, Michael and De Giovannini, Umberto and H\"ubener, Hannes and Rubio, Angel and Sentef, Michael A. and Devereaux, Thomas P. and Werner, Philipp},
  journal = {Phys. Rev. X},
  volume = {10},
  issue = {4},
  pages = {041013},
  numpages = {17},
  year = {2020},
  month = {Oct},
  publisher = {American Physical Society},
  doi = {10.1103/PhysRevX.10.041013},
}

@article{Peierls1933,
  title={Zur Theorie des Diamagnetismus von Leitungselektronen},
  author={Peierls, Rudolf},
  journal={Z. Phys.},
  volume={80},
  number={11--12},
  pages={763--791},
  year={1933},
  doi={10.1007/BF01342591}
}

@article{Shirley1965,
  title={Solution of the Schr{\"o}dinger Equation with a Hamiltonian Periodic in Time},
  author={Shirley, Jon H.},
  journal={Phys. Rev.},
  volume={138},
  number={4B},
  pages={B979--B987},
  year={1965},
  doi={10.1103/PhysRev.138.B979}
}

@article{Sambe1973,
  title={Steady States and Quasienergies of a Quantum-Mechanical System in an Oscillating Field},
  author={Sambe, Hideo},
  journal={Phys. Rev. A},
  volume={7},
  number={6},
  pages={2203--2213},
  year={1973},
  doi={10.1103/PhysRevA.7.2203}
}

@book{Barron2009,
  title={Molecular Light Scattering and Optical Activity},
  author={Barron, Laurence D.},
  edition={2},
  year={2009},
  publisher={Cambridge University Press},
  address={Cambridge}
}

@article{Mun2020,
  title={Electromagnetic chirality: from fundamentals to nontraditional chiroptical phenomena},
  author={Mun, Jungho and Kim, Minkyung and Yang, Younghwan and Badloe, Trevon and Ni, Jincheng and Chen, Yang and Qiu, Cheng-Wei and Rho, Junsuk},
  journal={Light: Science \& Applications},
  volume={9},
  number={1},
  pages={139},
  year={2020},
  publisher={Nature Publishing Group UK London}
}

@article{Lininger2023,
  title={Chirality in Light--Matter Interaction},
  author={Lininger, Andrew and Palermo, Giovanna and Guglielmelli, Alexa and Nicoletta, Giuseppe and Goel, Madhav and Hinczewski, Michael and Strangi, Giuseppe},
  journal={Adv. Mater.},
  volume={35},
  number={34},
  pages={2107325},
  year={2023},
  doi={10.1002/adma.202107325}
}

@ARTICLE{Wang2023,
  title     = "Optical activity of solids from first principles",
  author    = "Wang, Xiaoming and Yan, Yanfa",
  journal   = "Phys. Rev. B",
  publisher = "American Physical Society (APS)",
  volume    =  107,
  number    =  4,
  pages     =  045201,
  month     =  jan,
  year      =  2023,
}

@ARTICLE{Ocana2023,
  title={Multipole theory of optical spatial dispersion in crystals},
  author={Pozo Oca{\~n}a, {\'O}scar and Souza, Ivo},
  journal={SciPost Phys.},
  volume={14},
  number={5},
  pages={118},
  year={2023}
}

@ARTICLE{Zabalo2023,
  title     = "Natural optical activity from density-functional perturbation
               theory",
  author    = "Zabalo, Asier and Stengel, Massimiliano",
  journal   = "Phys. Rev. Lett.",
  publisher = "American Physical Society (APS)",
  volume    =  131,
  number    =  8,
  pages     =  086902,
  month     =  aug,
  year      =  2023,
}

@ARTICLE{Multunas2023,
  title     = "Circular dichroism of crystals from first principles",
  author    = "Multunas, Christian and Grieder, Andrew and Xu, Junqing and Ping,
               Yuan and Sundararaman, Ravishankar",
  journal   = "Phys. Rev. Mater.",
  publisher = "American Physical Society (APS)",
  volume    =  7,
  number    =  12,
  pages     =  123801,
  month     =  dec,
  year      =  2023,
}

@ARTICLE{Urru2025,
  title     = "Optical spatial dispersion via {Wannier} interpolation",
  author    = {Urru, Andrea and Souza, Ivo and Oca{\~n}a, {\'O}scar Pozo and Tsirkin, Stepan S and Vanderbilt, David},
  journal   = "Phys. Rev. B",
  publisher = "American Physical Society (APS)",
  volume    =  112,
  number    =  4,
  pages     =  045201,
  month     =  jul,
  year      =  2025,
}

@article{Petralli1993,
  title={Circular dichroism spectroscopy at interfaces: a surface second harmonic generation study},
  author={Petralli-Mallow, T and Wong, TM and Byers, JD and Yee, HI and Hicks, JM},
  journal={J. Phys. Chem.},
  volume={97},
  number={7},
  pages={1383--1388},
  year={1993},
  publisher={ACS Publications}
}

@article{Byers1994,
  title={Second-harmonic generation circular-dichroism spectroscopy from chiral monolayers},
  author={Byers, JD and Yee, HI and Petralli-Mallow, T and Hicks, JM},
  journal={Phys. Rev. B},
  volume={49},
  number={20},
  pages={14643},
  year={1994},
  publisher={APS}
}

@ARTICLE{Rodrigues2014,
  title     = "Nonlinear imaging and spectroscopy of chiral metamaterials",
  author    = "Rodrigues, Sean P and Lan, Shoufeng and Kang, Lei and Cui,
               Yonghao and Cai, Wenshan",
  journal   = "Adv. Mater.",
  publisher = "Wiley",
  volume    =  26,
  number    =  35,
  pages     = "6157--6162",
  month     =  sep,
  year      =  2014,
}

@article{Wang2016,
  title={Optical chiral metamaterials: a review of the fundamentals, fabrication methods and applications},
  author={Wang, Zuojia and Cheng, Feng and Winsor, Thomas and Liu, Yongmin},
  journal={Nanotechnology},
  volume={27},
  number={41},
  pages={412001},
  year={2016},
  publisher={IOP Publishing}
}

@article{Chen2016,
  title={Giant nonlinear optical activity of achiral origin in planar metasurfaces with quadratic and cubic nonlinearities},
  author={Chen, Shumei and Zeuner, Franziska and Weismann, Martin and Reineke, Bernhard and Li, Guixin and Valev, Ventsislav Kolev and Cheah, Kok Wai and Panoiu, Nicolae Coriolan and Zentgraf, Thomas and Zhang, Shuang},
  journal={Adv. Mater.},
  volume={28},
  number={15},
  pages={2992--2999},
  year={2016}
}

@article{Li2017,
  title={Nonlinear photonic metasurfaces},
  author={Li, Guixin and Zhang, Shuang and Zentgraf, Thomas},
  journal={Nat. Rev. Mater.},
  volume={2},
  number={5},
  pages={17010},
  year={2017},
  publisher={Nature Publishing Group}
}

@article{Tang2020,
  title={Nano-Kirigami metasurface with giant nonlinear optical circular dichroism},
  author={Tang, Yutao and Liu, Zhiguang and Deng, Junhong and Li, Kingfai and Li, Jiafang and Li, Guixin},
  journal={Laser Photonics Rev.},
  volume={14},
  number={7},
  pages={2000085},
  year={2020},
  publisher={Wiley Online Library}
}

@article{Kim2020,
  title={Giant nonlinear circular dichroism from intersubband polaritonic metasurfaces},
  author={Kim, Daeik and Yu, Jaeyeon and Hwang, Inyong and Park, Seongjin and Demmerle, Frederic and Boehm, Gerhard and Amann, Markus-Christian and Belkin, Mikhail A and Lee, Jongwon},
  journal={Nano Lett.},
  volume={20},
  number={11},
  pages={8032--8039},
  year={2020},
  publisher={ACS Publications}
}

@article{Frizyuk2021,
  title={Nonlinear circular dichroism in {Mie}-resonant nanoparticle dimers},
  author={Frizyuk, Kristina and Melik-Gaykazyan, Elizaveta and Choi, Jae-Hyuck and Petrov, Mihail I and Park, Hong-Gyu and Kivshar, Yuri},
  journal={Nano Lett.},
  volume={21},
  number={10},
  pages={4381--4387},
  year={2021},
  publisher={ACS Publications}
}

@article{Gandolfi2021,
  title={Near-unity third-harmonic circular dichroism driven by a quasibound state in the continuum in asymmetric silicon metasurfaces},
  author={Gandolfi, Marco and Tognazzi, Andrea and Rocco, Davide and De Angelis, Costantino and Carletti, Luca},
  journal={Phys. Rev. A},
  volume={104},
  number={2},
  pages={023524},
  year={2021},
  publisher={APS}
}

@article{Nikitina2023,
  title={Nonlinear circular dichroism in achiral dielectric nanoparticles},
  author={Nikitina, Anastasia and Nikolaeva, Anna and Frizyuk, Kristina},
  journal={Phys. Rev. B},
  volume={107},
  number={4},
  pages={L041405},
  year={2023},
  publisher={APS}
}

@article{Toftul2024,
  title = {Chiral Dichroism in Resonant Metasurfaces with Monoclinic Lattices},
  author = {Toftul, Ivan and Tonkaev, Pavel and Koshelev, Kirill and Lai, Fangxing and Song, Qinghai and Gorkunov, Maxim and Kivshar, Yuri},
  journal = {Phys. Rev. Lett.},
  volume = {133},
  issue = {21},
  pages = {216901},
  numpages = {6},
  year = {2024},
  month = {Nov},
  publisher = {American Physical Society},
  doi = {10.1103/PhysRevLett.133.216901}
}

@ARTICLE{Lovesey2019,
  title={Circular dichroism of second harmonic generation response},
  author={Lovesey, SW and van der Laan, G},
  journal={Phys. Rev. B},
  volume={100},
  number={24},
  pages={245112},
  year={2019},
  publisher={APS}
}

@ARTICLE{Chen2020,
  title={Probing structural chirality of crystals using high-order harmonic generation in solids},
  author={Chen, Zi-Yu and Qin, Rui},
  journal={Phys. Rev. A},
  volume={101},
  number={5},
  pages={053423},
  year={2020},
  publisher={APS}
}

@book{Bradley1972,
  author    = "C. J. Bradley and A. P. Cracknell",
  title     = "The Mathematical Theory of Symmetry in Solids",
  year      = "1972",
  publisher = "Clarendon",
  address   = "Oxford"
}

@book{Bloembergen1996,
  author    = "N. Bloembergen",
  title     = "Nonlinear Optics",
  edition   = "4",
  year      = "1996",
  publisher = "World Scientific",
  address   = "Singapore"
}

@book{Boyd2008,
  author    = "R. W. Boyd",
  title     = "Nonlinear Optics",
  edition   = "3",
  year      = "2008",
  publisher = "Academic Press",
  address   = "London"
}

@article{Morimoto2016,
  title     = "Topological Nature of Nonlinear Optical Effects in Solids",
  author    = "T. Morimoto and N. Nagaosa",
  journal   = "Sci. Adv.",
  volume    = "2",
  number    = "5",
  pages     = "e1501524",
  year      = "2016",
  publisher = "AAAS"
}

@article{Lv2017,
  author    = "B. Q. Lv and Z.-L. Feng and Q.-N. Xu and X. Gao and J.-Z. Ma and L.-Y. Kong and P. Richard and Y.-B. Huang and V. N. Strocov and C. Fang and H.-M. Weng and Y.-G. Shi and T. Qian and H. Ding",
  title     = "Observation of three-component fermions in the topological semimetal molybdenum phosphide",
  journal   = "Nature",
  volume    = "546",
  number    = "7660",
  pages     = "627--631",
  year      = "2017",
  doi       = "10.1038/nature22390"
}

@ARTICLE{Tang2017,
  title    = "Multiple Types of Topological Fermions in Transition Metal Silicides",
  author   = "Tang, Peizhe and Zhou, Quan and Zhang, Shou-Cheng",
  journal  = "Phys. Rev. Lett.",
  volume   =  119,
  number   =  20,
  pages    =  206402,
  month    =  nov,
  year     =  2017,
}

@ARTICLE{Chang2017,
  title    = "Unconventional Chiral Fermions and Large Topological {Fermi} Arcs in {RhSi}",
  author   = "Chang, Guoqing and Xu, Su-Yang and Wieder, Benjamin J and Sanchez,
              Daniel S and Huang, Shin-Ming and Belopolski, Ilya and Chang,
              Tay-Rong and Zhang, Songtian and Bansil, Arun and Lin, Hsin and
              Hasan, M Zahid",
  journal  = "Phys. Rev. Lett.",
  volume   =  119,
  number   =  20,
  pages    =  206401,
  month    =  nov,
  year     =  2017,
}

@article{Pshenay-Severin2018,
  title     = "Band Structure and Unconventional Electronic Topology of {CoSi}",
  author    = "Pshenay-Severin, D. A. and Ivanov, Y. V. and Burkov, A. A. and Burkov, A. T.",
  journal   = "J. Phys. Condens. Matter",
  volume    = "30",
  number    = "13",
  pages     = "135501",
  year      = "2018",
  month     = "Apr",
  publisher = "IOP Publishing"
}

@ARTICLE{Rao2019,
  title    = "Observation of unconventional chiral fermions with long {Fermi} arcs
              in {CoSi}",
  author   = "Rao, Zhicheng and Li, Hang and Zhang, Tiantian and Tian, Shangjie
              and Li, Chenghe and Fu, Binbin and Tang, Cenyao and Wang, Le and
              Li, Zhilin and Fan, Wenhui and Li, Jiajun and Huang, Yaobo and
              Liu, Zhehong and Long, Youwen and Fang, Chen and Weng, Hongming
              and Shi, Youguo and Lei, Hechang and Sun, Yujie and Qian, Tian and
              Ding, Hong",
  journal  = "Nature",
  volume   =  567,
  number   =  7749,
  pages    = "496--499",
  month    =  mar,
  year     =  2019,
}

@ARTICLE{Takane2019,
  title    = "Observation of Chiral Fermions with a Large Topological Charge and
              Associated Fermi-Arc Surface States in {CoSi}",
  author   = "Takane, Daichi and Wang, Zhiwei and Souma, Seigo and Nakayama,
              Kosuke and Nakamura, Takechika and Oinuma, Hikaru and Nakata, Yuki
              and Iwasawa, Hideaki and Cacho, Cephise and Kim, Timur and Horiba,
              Koji and Kumigashira, Hiroshi and Takahashi, Takashi and Ando,
              Yoichi and Sato, Takafumi",
  journal  = "Phys. Rev. Lett.",
  volume   =  122,
  number   =  7,
  pages    =  076402,
  month    =  feb,
  year     =  2019,
}

@ARTICLE{Sanchez2019,
  title    = "Topological chiral crystals with helicoid-arc quantum states",
  author   = "Sanchez, Daniel S and Belopolski, Ilya and Cochran, Tyler A and
              Xu, Xitong and Yin, Jia-Xin and Chang, Guoqing and Xie, Weiwei and
              Manna, Kaustuv and Süß, Vicky and Huang, Cheng-Yi and Alidoust,
              Nasser and Multer, Daniel and Zhang, Songtian S and Shumiya, Nana
              and Wang, Xirui and Wang, Guang-Qiang and Chang, Tay-Rong and
              Felser, Claudia and Xu, Su-Yang and Jia, Shuang and Lin, Hsin and
              Hasan, M Zahid",
  journal  = "Nature",
  volume   =  567,
  number   =  7749,
  pages    = "500--505",
  month    =  mar,
  year     =  2019,
}

@article{Schroter2019,
  title     = "Chiral Topological Semimetal with Multifold Band Crossings and Long {Fermi} Arcs",
  author    = "Schr{\"o}ter, Niels B. M. and Pei, Ding and Vergniory, Maia G. and Sun, Yan and Manna, Kaustuv and de Juan, Fernando and Krieger, Jonas A. and S{\"u}ss, Vicky and Schmidt, Marcus and Dudin, Pavel and Bradlyn, Barry and Kim, Timur K. and Schmitt, Thorsten and Cacho, Cephise and Felser, Claudia and Strocov, Vladimir N. and Chen, Yulin",
  journal   = "Nat. Phys.",
  volume    = "15",
  number    = "8",
  pages     = "759--765",
  year      = "2019",
  publisher = "Nature Publishing Group",
  doi       = "10.1038/s41567-019-0511-y"
}

@ARTICLE{Li2019,
  title     = "Chiral fermion reversal in chiral crystals",
  author    = "Li, Hang and Xu, Sheng and Rao, Zhi-Cheng and Zhou, Li-Qin and
               Wang, Zhi-Jun and Zhou, Shi-Ming and Tian, Shang-Jie and Gao,
               Shun-Ye and Li, Jia-Jun and Huang, Yao-Bo and Lei, He-Chang and
               Weng, Hong-Ming and Sun, Yu-Jie and Xia, Tian-Long and Qian, Tian
               and Ding, Hong",
  journal   = "Nat. Commun.",
  publisher = "Springer Science and Business Media LLC",
  volume    =  10,
  number    =  1,
  pages     =  5505,
  month     =  dec,
  year      =  2019
}

@ARTICLE{Ahn2023,
  title     = "Topological Circular Dichroism in Chiral Multifold Semimetals",
  author    = "Ahn, Junyeong and Ghosh, Barun",
  journal   = "Phys. Rev. Lett.",
  publisher = "American Physical Society",
  volume    =  131,
  number    =  11,
  pages     =  116603,
  month     =  sep,
  year      =  2023,
}

@ARTICLE{Fan2024,
  title={Chiral {Floquet} engineering on topological fermions in chiral crystals},
  author={Fan, Benshu and Duan, Wenhui and Rubio, Angel and Tang, Peizhe},
  journal={npj Quantum Mater.},
  volume={9},
  number={1},
  pages={101},
  year={2024},
  publisher={Nature Publishing Group UK London}
}

@article{Satow2025,
  title = {Symmetry-adapted models for multifold fermions with spin-orbit coupling},
  author = {Satow, Koki and Yamakage, Ai},
  journal = {Phys. Rev. B},
  volume = {112},
  issue = {19},
  pages = {195206},
  numpages = {13},
  year = {2025},
  month = {Nov},
  publisher = {American Physical Society},
  doi = {10.1103/m8c6-yvt3},
  url = {https://link.aps.org/doi/10.1103/m8c6-yvt3}
}

@ARTICLE{Flicker2018,
  title     = "Chiral optical response of multifold fermions",
  author    = "Flicker, Felix and de Juan, Fernando and Bradlyn, Barry and
               Morimoto, Takahiro and Vergniory, Maia G and Grushin, Adolfo G",
  journal   = "Phys. Rev. B",
  publisher = "American Physical Society",
  volume    =  98,
  number    =  15,
  pages     =  155145,
  month     =  oct,
  year      =  2018
}

@article{Ni2020,
  title     = "Linear and Nonlinear Optical Responses in the Chiral Multifold Semimetal {RhSi}",
  author    = "Ni, Zhuoliang and Xu, B. and S{\'a}nchez-Mart{\'\i}nez, M-{\'A}. and Zhang, Y. and Manna, K. and Bernhard, C. and Venderbos, J. W. F. and de Juan, F. and Felser, C. and Grushin, Adolfo G. and Liang Wu",
  journal   = "npj Quantum Mater.",
  volume    = "5",
  number    = "1",
  pages     = "96",
  year      = "2020",
  publisher = "Nature Publishing Group"
}

@article{Rees2020,
  title     = "Helicity-Dependent Photocurrents in the Chiral Weyl Semimetal {RhSi}",
  author    = "Rees, Dylan and Manna, Kaustuv and Lu, Baozhu and Morimoto, Takahiro and Borrmann, Horst and Felser, Claudia and Moore, J. E. and Torchinsky, Darius H. and Orenstein, J.",
  journal   = "Sci. Adv.",
  volume    = "6",
  number    = "29",
  pages     = "eaba0509",
  year      = "2020",
  publisher = "American Association for the Advancement of Science"
}

@article{Ni2021,
  title     = "Giant Topological Longitudinal Circular Photo-Galvanic Effect in the Chiral Multifold Semimetal {CoSi}",
  author    = "Ni, Zhuoliang and Wang, K. and Zhang, Y. and Pozo, O. and Xu, B. and Han, X. and Manna, K. and Paglione, J. and Felser, C. and Grushin, Adolfo G. and F. de Juan and E. J. Mele and Liang Wu",
  journal   = "Nat. Commun.",
  volume    = "12",
  number    = "1",
  pages     = "154",
  year      = "2021",
  publisher = "Nature Publishing Group"
}

@article{Xu2020,
  title     = "Optical Signatures of Multifold Fermions in the Chiral Topological Semimetal {CoSi}",
  author    = "Xu, Bing and Fang, Zhenyao and S{\'a}nchez-Mart{\'\i}nez, Miguel-{\'A}ngel and Venderbos, Jorn W. F. and Ni, Zhuoliang and Qiu, Tian and Manna, Kaustuv and Wang, Kefeng and Paglione, Johnpierre and Bernhard, Christian and Felser, Claudia and Mele, E. J. and Grushin, Adolfo G. and Rappe, Andrew M. and Wu, Liang",
  journal   = "Proc. Natl. Acad. Sci. U.S.A.",
  volume    = "117",
  number    = "44",
  pages     = "27104--27110",
  year      = "2020",
  publisher = "National Academy of Sciences"
}

@ARTICLE{Lu2022,
  title={Second-harmonic generation in the topological multifold semimetal {RhSi}},
  author={Lu, Baozhu and Sayyad, Sharareh and S{\'a}nchez-Mart{\'\i}nez, Miguel {\'A}ngel and Manna, Kaustuv and Felser, Claudia and Grushin, Adolfo G and Torchinsky, Darius H},
  journal={Phys. Rev. Res.},
  volume={4},
  number={2},
  pages={L022022},
  year={2022},
  publisher={APS}
}

@ARTICLE{Tsuji2008,
  title     = "Correlated electron systems periodically driven out of
               equilibrium: {FloquetDMFT} formalism",
  author    = "Tsuji, Naoto and Oka, Takashi and Aoki, Hideo",
  journal   = "Phys. Rev. B",
  publisher = "American Physical Society",
  volume    =  78,
  number    =  23,
  pages     =  235124,
  month     =  dec,
  year      =  2008
}

@ARTICLE{Tsuji2009,
  title    = "Nonequilibrium steady state of photoexcited correlated electrons
              in the presence of dissipation",
  author   = "Tsuji, Naoto and Oka, Takashi and Aoki, Hideo",
  journal  = "Phys. Rev. Lett.",
  volume   =  103,
  number   =  4,
  pages    =  047403,
  month    =  jul,
  year     =  2009,
}

@ARTICLE{Aoki2014,
  title     = "Nonequilibrium dynamical mean-field theory and its applications",
  author    = "Aoki, Hideo and Tsuji, Naoto and Eckstein, Martin and Kollar,
               Marcus and Oka, Takashi and Werner, Philipp",
  journal   = "Rev. Mod. Phys.",
  publisher = "American Physical Society",
  volume    =  86,
  number    =  2,
  pages     = "779--837",
  month     =  jun,
  year      =  2014
}

@article{Simon1968,
  title = {Second-Harmonic Light Generation in Crystals with Natural Optical Activity},
  author = {Simon, H. J. and Bloembergen, N.},
  journal = {Phys. Rev.},
  volume = {171},
  issue = {3},
  pages = {1104--1114},
  numpages = {0},
  year = {1968},
  month = {Jul},
  publisher = {American Physical Society},
  doi = {10.1103/PhysRev.171.1104}
}

@article{Tang1971,
  title = {Selection Rules for Circularly Polarized Waves in Nonlinear Optics},
  author = {Tang, C. L. and Rabin, Herbert},
  journal = {Phys. Rev. B},
  volume = {3},
  issue = {12},
  pages = {4025--4034},
  numpages = {0},
  year = {1971},
  month = {Jun},
  publisher = {American Physical Society},
  doi = {10.1103/PhysRevB.3.4025}
}

@article{Alon1998,
  title     = "Selection Rules for the High Harmonic Generation Spectra",
  author    = "Alon, Ofir E. and Averbukh, Vitali and Moiseyev, Nimrod",
  journal   = "Phys. Rev. Lett.",
  volume    = "80",
  number    = "17",
  pages     = "3743--3746",
  year      = "1998",
  publisher = "American Physical Society"
}

@article{Saito2017,
  title     = "Observation of Selection Rules for Circularly Polarized Fields in High-Harmonic Generation from a Crystalline Solid",
  author    = "Saito, Nariyuki and Xia, Peiyu and Lu, Faming and Kanai, Teruto and Itatani, Jiro and Ishii, Nobuhisa",
  journal   = "Optica",
  volume    = "4",
  number    = "11",
  pages     = "1333--1336",
  year      = "2017",
  month     = "Nov",
  publisher = "Optica Publishing Group"
}

@ARTICLE{Neufeld2019,
  title     = "Floquet group theory and its application to selection rules in
               harmonic generation",
  author    = "Neufeld, Ofer and Podolsky, Daniel and Cohen, Oren",
  journal   = "Nat. Commun.",
  publisher = "Springer Science and Business Media LLC",
  volume    =  10,
  number    =  1,
  pages     =  405,
  month     =  jan,
  year      =  2019,
}

@article{Ikeda2019,
  title     = "High-Harmonic Generation by Electric Polarization, Spin Current, and Magnetization",
  author    = "Ikeda, Tatsuhiko N. and Sato, Masahiro",
  journal   = "Phys. Rev. B",
  volume    = "100",
  number    = "21",
  pages     = "214424",
  year      = "2019",
  publisher = "American Physical Society"
}

@article{Ikeda2020,
  title     = "High-Order Nonlinear Optical Response of a Twisted Bilayer Graphene",
  author    = "Ikeda, Tatsuhiko N.",
  journal   = "Phys. Rev. Res.",
  volume    = "2",
  number    = "3",
  pages     = "032015",
  year      = "2020",
  publisher = "American Physical Society"
}

@article{Kanega2021,
  title     = "Linear and Nonlinear Optical Responses in {Kitaev} Spin Liquids",
  author    = "Kanega, Minoru and Ikeda, Tatsuhiko N. and Sato, Masahiro",
  journal   = "Phys. Rev. Res.",
  volume    = "3",
  number    = "3",
  pages     = "L032024",
  year      = "2021",
  publisher = "American Physical Society"
}

@article{Kanega2024,
  title     = "High-Harmonic Generation in Graphene under the Application of a DC Electric Current: From Perturbative to Nonperturbative Regime",
  author    = "Kanega, Minoru and Sato, Masahiro",
  journal   = "Phys. Rev. B",
  volume    = "110",
  number    = "3",
  pages     = "035303",
  year      = "2024",
  publisher = "American Physical Society"
}

@ARTICLE{Ominato2025,
  title     = "Theory of photocurrent and high-harmonic generation with chiral
               fermions",
  author    = "Ominato, Yuya and Mochizuki, Masahito",
  journal   = "Phys. Rev. Res.",
  publisher = "American Physical Society",
  volume    =  7,
  pages     =  023218,
  month     =  jun,
  year      =  2025,
}

\end{document}